\documentclass[10pt]{amsart}
\usepackage{backref}

     \usepackage[german,english]{babel}
     \usepackage{enumerate}
     \usepackage{exscale}
     \usepackage{a4}
     \usepackage{amsmath,amsfonts,amssymb}
     \newcommand{\be}{\begin{equation}}
     \newcommand{\ee}{\end{equation}}
     \newcommand{\bea}{\begin{eqnarray*}}
     \newcommand{\eea}{\end{eqnarray*}}
     \newcommand{\beq}{\begin{eqnarray}}
     \newcommand{\eeq}{\end{eqnarray}}
     \newcommand{\CC}{\mathbb{C}}
     \newcommand{\EE}{\mathbb{E}}
     
     \newcommand{\NN}{\mathbb{N}}
     \newcommand{\PP}{\mathbb{P}}
     \newcommand{\QQ}{\mathbb{Q}}
     \newcommand{\RR}{\mathbb{R}}
     \newcommand{\ZZ}{\mathbb{Z}}
     \newcommand{\cA}{\mathcal{A}}
     
     \newcommand{\cD}{\mathcal{D}}
     \newcommand{\cF}{\mathcal{F}}

     \newcommand{\cN}{\mathcal{N}}
     \newcommand{\cR}{\mathcal{R}}

     \renewcommand{\r}{\right}
     \renewcommand{\l}{\left}
     \newcommand{\la}{\langle}
     \newcommand{\ra}{\rangle}
     \newcommand{\Ra}{\Rightarrow}
     \newcommand{\iznad}[2]{\genfrac{}{}{0pt}{}{#1}{#2}}
     \newcommand{\nr}[1]{\vert #1\vert}
     
     \newcommand{\supp}{\mathop{\mathrm{supp}}}
     
     \newcommand{\Tr}{\mathop{\mathrm{Tr}}}
     \newcommand{\dist}{\mathrm{dist}}
     \newcommand{\proj}{\mathrm{proj}}
     \newcommand{\kom}[1]{}
     
     \DeclareMathOperator*{\ran}{\mathrm{ran}}
     
     \newtheorem{thm}{Theorem}[section]
     \newtheorem{lem}[thm]{Lemma}
     \newtheorem{prp}[thm]{Proposition}
     \newtheorem{cor}[thm]{Corollary}
     \theoremstyle{definition}
     \newtheorem{dfn}[thm]{Definition}
     
     \theoremstyle{remark}
     \newtheorem*{rem}{Remark}

\newcount\minute    
\newcount\hour      
\newcount\hourMins  
\def\now%
{
  \minute=\time    
  \hour=\time \divide \hour by 60 
  \hourMins=\hour \multiply\hourMins by 60
  \advance\minute by -\hourMins 
  \zeroPadTwo{\the\hour}:\zeroPadTwo{\the\minute}%
}
\def\zeroPadTwo#1%
{
  \ifnum #1<10 0\fi    
  #1
}

\begin{document}

\title{Spectral Analysis of Percolation Hamiltonians}

\author{Ivan Veseli\'c}
\address{Fakult\"at f\"ur Mathematik, D-09107 TU Chemnitz }
\email{ivan.veselic@mathematik.tu-chemnitz.de}
\urladdr{www.tu-chemnitz.de/mathematik/schroedinger/index.php}


\begin{abstract}
We study the family of Hamiltonians which corresponds to the adjacency operators on a percolation graph.
We characterise the set of energies which are almost surely eigenvalues with finitely supported eigenfunctions.
This set of energies is a dense subset of the algebraic integers. The integrated density of states has discontinuities precisely at this set of energies. We show that the convergence of the integrated densities of states of finite box Hamiltonians to the one on the whole space holds even at the points of discontinuity. For this we use an equicontinuity-from-the-right argument. The same statements hold for the restriction of the Hamiltonian to the infinite cluster. In this case we prove that the integrated density of states can be constructed using local data only.
Finally we study some mixed Anderson-Quantum percolation models and establish results in the spirit of Wegner, and Delyon and Souillard.
\end{abstract}

\thanks{See also arXiv.org/math-ph/0405006. To appear in a slightly different version in \emph{Math.Ann.} with DOI 10.1007/s00208-004-0610-6.}
\maketitle

\section{Introduction: The Quantum percolation model (QPM)}
The present paper is devoted to the spectral analysis of the percolation Hamiltonian. 
It is the family of adjacency operators associated to realisations of percolation sub-graphs of the $d$-dimensional lattice. 
For simplicity we restrict ourselves in the present paper to site percolation, although
most of the results are valid also for bond percolation. This will be discussed elsewhere in more detail.

Unlike other random lattice Hamiltonians, the quantum percolation model has finitely supported eigenstates.
Indeed, this property was the first cause of interest in the model, cf.~\cite{deGennesLM-59a}.
Another feature, which sets the quantum percolation model apart from other random Hamiltonians, 
is the existence of a large set of discontinuity points of its integrated density of states (IDS).
Nevertheless, we are able to prove the convergence of the finite volume approximants to the IDS at \emph{all} energies. 

In the mathematical physics literature the continuity of the integrated density of states has been proven for several types of random Schr\"odinger operators. The considered models act as differential operators on $L^2(\RR^d)$ or as difference operators on $\ell^2(\ZZ^d)$. On the other hand, for some models, like the Bernoulli-alloy type model, the continuity is still an open question. By means of contrast, the QPM may provide understanding what the relevant mechanisms are which cause the continuity. In the literature on equivariant manifolds and graphs and of geometric $L^2$-invariants the analogue of the IDS is studied, too. It has been noticed that its discontinuities contain geometric information of the underlying space, see \cite{Lueck-02} and the references cited there. 
The discontinuities of the IDS of certain different random Hamiltonians have been recently studied in \cite{KlassertLS-03} and \cite{KostrykinS-04}. The first paper is devoted to tiling Hamiltonians and the second to the random necklace model.

\medskip

The QPM was introduced by de Gennes, Lafore, and Millot in 1959 \cite{deGennesLM-59a,deGennesLM-59b}.
There it was considered as the Hamiltonian of a binary solid solution. De Gennes et al.~showed that the spectrum of the percolation Hamiltonian is pure point if the fraction $p$ of active sites is below a critical value $p_c$.  
Indeed, this value is the well known critical probability of percolation theory: If $p<p_c$ no infinite active cluster exists almost surely, and for $p>p_c$ it almost surely does.
If the concentration of active sites is above the critical value one speaks of the \emph{percolating regime}.
For this regime  de Gennes et al.~argued that the spectrum contains a continuous part. This statement was not verified by rigorous methods. Its proof would provide an example of \emph{Anderson de-localisation}, see the paragraph below for more details.
In \cite{KirkpatrickE-72} Kirkpatrick and Eggarter observed that there exist eigenfunctions of the adjacency operator whose support is finite and contained in the infinite percolation cluster. Obviously this statement is only interesting in the percolating regime.
It is easy to find such eigenfunctions corresponding to the eigenvalue zero (the midpoint of the spectrum). To show the existence of non-zero eigenvalues Kirkpatrick and Eggarter constructed examples with an axis of symmetry. Their idea resembles the ``mirror charges" construction in electro statics. 

In \cite{ChayesCFST-86} Chayes, Chayes, Franz, Sethna, and Trugman refined the mirror charge idea. They presented arguments that the eigenvalues of finitely supported states form a dense set in the spectrum, and that the IDS is discontinuous at these energies. We give a complete mathematical proof of these facts 
and moreover show that \emph{any} energy occurring in the spectrum of the adjacency operator of a finite cluster also occurs with positive probability as an eigenvalue (with finitely supported state) on the infinite cluster. Furthermore, we show that the set of these energies coincides with the set of discontinuities of the IDS. 
Actually, the argument showing the equality of these two sets of energies is by no means restricted to percolation 
Hamiltonians, but applies to much more general random operators on graphs. This will be discussed elsewhere.
\medskip

Anderson localisation and delocalisation go beyond the scope of this paper. Nevertheless, for completeness sake 
we briefly discuss these topics and give references to the physics literature where they were studied numerically for the QPM. In the fifties Anderson \cite{Anderson-58} argued on physical grounds that certain lattice Hamiltonians describing the motion of a single electron in a disordered environment should exhibit pure point spectrum. (The model he studied is the same as $H_\omega$ defined in \S \ref{s-results}, with the assumption that the random variables $q_k, k\in \ZZ^d$ are uniformly distributed on a finite interval.)
Later it has been proved rigorously that the spectrum of Andersons's model near its boundaries consists of a dense set of eigenvalues and that the continuous spectral component is absent, see the expositions in \cite{CarmonaL-90,PasturF-92,Stollmann-01}. Moreover, all the corresponding eigenfunctions decay exponentially.
This phenomenon is called \emph{Anderson localisation}. The exponential decay rate is characterised by the so called Lyapunov exponent, or its inverse, the localisation length. 
It is conjectured, but not proven that in a certain energy/disorder regime the Anderson model should exhibit \emph{delocalisation}, i.e.~purely continuous spectrum. 

Physical intuition suggests that the QPM should also exhibit Anderson localisation in energy regions near spectral boundaries. If this holds, it is 
a priori not clear whether the exponentially localised eigenfunctions could induce a discontinuity of the IDS.
We show that this is not the case.

A series of papers is devoted to the numerical analysis of Anderson localisation for the QPM, e.g~\cite{ShapirAH-82,KantelhardtB-97,KantelhardtB-98,KantelhardtB-98b,KantelhardtB-02}. These studies include a discussion of the similarities and differences of localisation occurring in quantum percolation and Anderson models. 
In particular the numerical analysis 
done in \cite{KantelhardtB-02} indicates that the localisation length for two dimensional systems and energies in the middle of the band behaves differently for the Anderson model and for the QPM. In the later case there is a well defined  localisation length (independent of the distance from the localisation centre), supporting the picture that the eigenstates decay exponentially. On the other hand, if one tries to define 
a localisation length for the Anderson model, it turns out that this quantity grows  logarithmically as a function of the distance  from the localisation centre. This indicates that the eigenfunction decay is slower than exponential. In \cite{ShapirAH-82} quantum percolation in the percolating regime $p>p_c$ is studied. Indications are found that in dimensions greater than two extended states (i.e.~continuous spectrum) emerge for $p>p_q$, where $p_q$ is a threshold probability strictly larger than $p_c$. 
\medskip

The proofs in the present paper draw on ideas both from the random Schr\"odinger operator and the $L^2$-invariants   literature. For instance, Delyon and Souillard in \cite{DelyonS-84} used the unique continuation property to prove the continuity of the IDS of the Anderson model. In \cite{CraigS-83a} Craig and Simon obtained a better result, namely the log-H\"older continuity of the IDS. While the techniques of these two papers are not applicable in the present context, the unique continuation property and the log-H\"older continuity do play a prominent role. Their relevance has been realised in the geometry literature too, see e.g.~the papers of L\"uck \cite{Lueck-94c}, Farber \cite{Farber-98} and Dodziuk, Mathai and Yates \cite{DodziukMY,MathaiY-02}. They are devoted to equivariant operators on graphs and manifolds. In particular, these operators are non-random. While these papers show that the discontinuities of the IDS have to be a subset of a certain algebraic set, they do not prove the existence of discontinuities and a characterisation of this set. In the present setting the ideas developed in papers on the theory of disordered systems and on the theory of geometric invariants turn out to be complementary: while the former ones are used to prove the existence of discontinuities and characterise their set, the latter ones allow one to prove the convergence of the approximations of the IDS even at energis where the IDS jumps.

In this paper we study Hamiltonians corresponding to site percolation on the lattice $\ZZ^d$ and on general graphs with an amenable group action. For clarity sake we separate the discussion and focus first on $\ZZ^d$ and later on 
analyse the more general case of graphs.
The types of  Hamiltonians we consider are somewhat more general than the QPM and contain the Anderson model as special case as well. 
Our results apply also to the case where the probability space degenerates to a point, i.e.~when the considered operator is invariant under a group action. 
\medskip
 
Here ist the outline of the paper: 
The following section contains precise definitions and statements of the results in the case where the the Hamiltonian is defined on $\ZZ^d$. In particular, we define the IDS by an exhaustion procedure and analyse its continuity properties. Subsequently we discuss in \S~\ref{s-general} which of the results hold on more general graphs and for correlated random potentials. The proofs in the later sections are given for this more general situation. 
In Section \ref{s-LocDef} we give an alternative, local definition of the IDS on the infinite cluster. 
It is followed by a section discussing the unique continuation property and characterising the discontinuities of the IDS. 
Section \ref{s-lH} proves the log-H\"older continuity of the IDS at all algebraic numbers, 
and deduces the convergence of the finite volume approximations at all energies. Section \ref{s-Wegner} proves the local Lipschitz continuity of the IDS under local continuity requirements on the potential values. There one can find also a generalisation of an argument of Delyon and Suillard \cite{DelyonS-84}.

\subsection*{Acknowledgements}
It is a pleasure to thank J.~Dodziuk, D.~Katz, W.~Kirsch, and D.~Lenz for stimulating discussions and comments; the Deutsche Forschungsgemeinschaft for support through grants Ve 253/1-1 and /2-1; the anonymous referees for clarifying remarks; and B.~Simon for hospitality at CalTech. 

\section{Results}
\label{s-results}
Let us first define the model we are studying. In the present section we will restrict ourselves to Hamiltonians corresponding to independent, identically distributed (iid) percolation on $\ZZ^d$. Consider a collection of iid random variables $q_k\colon \Omega \to [0,\infty]$ indexed by $k \in \ZZ^d$.  
Here $\Omega$ is a probability space with probability measure $\PP$ and expectation $\EE$. 
Denote by $\mu$ be the Borel probability measure corresponding to the distribution of the random variable $q_0$. 
Then $\PP =\otimes_{\ZZ^d} \mu $. 
To each $\omega\in \Omega$ corresponds a 
function $q_\bullet (\omega) \colon \ZZ^d \to [0,\infty]$ which is called a
\emph{realisation} or \emph{configuration}.
It defines the  
subset $X(\omega):= \{k \mid q_k(\omega) < \infty \}$. Sometimes we will call $\omega\in \Omega$ itself a configuration.
The vertices in $X(\omega)$ are called \emph{active} or \emph{open} and the ones outside $X(\omega)$ \emph{closed}. To avoid trivialities, suppose $p:=\PP\{q_0(\omega)<\infty\}>0$. We could also allow that the  random variables take on negative values, if we impose an appropriate moment condition.

A \emph{finite hopping range operator} $H_0$ on 
$\ZZ^d$ is a bounded, linear map $H_0\colon \ell^2(\ZZ^d)\to \ell^2(\ZZ^d)$ such that there exists $ R\in\NN$ with
\begin{enumerate}[\rm(i)]
\label{i-fr}
\item $H_0(k,j) = H_0(j,k) \in \RR,$
\item $H_0( k+l,j+l) = H_0(k,j)$ for all  $l \in \ZZ^d$ and
\item $H_0(k,j)=0$ if $\|k-j\|_1\ge R$
\end{enumerate} 
for all $k,j \in \ZZ^d$. Here $H_0(k,j):= \la \delta_k, H_0 \delta_j\ra$ and $\delta_k\in \ell^2(\ZZ^d)$
is the function  taking the value $1$ at $k$ and $0$ elsewhere. 
It follows that there exists a constant $C$ with $|H_0(k,j)| \le C$ for all $k,j \in \ZZ^d$.
The smallest $R\in \NN$ for which (iii) holds is the \emph{finite hopping range} of $H_0$.
The most important example for $H_0$ is the \emph{adjacency operator}: $H_0(k,j)=1$ if $\|k-j\|_{1}=1$, and $0$ otherwise.

We think of $H_0$ as the kinetic energy of a Hamiltonian. The potential energy is given by the multiplication operator by $q(\omega)$ and the full \emph{random Hamiltonian} $H_\omega$ is the sum of the two energies. More precisely, for an $\omega \in \Omega$, define 
$\cD(H_\omega) := \ell^2(X(\omega))$ and 
\[
(H_\omega f)(k):= q_k(\omega)f(k) + \sum_{j\in X(\omega)} H(k,j)f(j)  \ \text{ for all } f \in \cD(H_\omega) 
\]
Thus $q_k(\omega)=\infty$ implies that functions in the operator domain $\cD(H_\omega)$ vanish at $k$. This is consistent with thinking that the potential value at $k$ is infinitely high.

For a subset $G \subset \ZZ^d$, an $\omega\in \Omega$ and a random Hamiltonian $H_\omega$ denote by
$H_\omega^G $ the restriction of $H_\omega$ to $\ell^2(G)$, in other words
$H_\omega^G(k,j) =  H_\omega(k,j)$ if $ k,j \in G $.
For finite $G$, the spectrum of $H_\omega^G$ consists of eigenvalues ${E}_1(H_\omega^G)\le {E}_2(H_\omega^G) \le \dots$, which we enumerate in increasing order including multiplicities.   
The normalized eigenvalue counting function of $H_\omega^G$ is defined as
\[
N(H_\omega^G,{E})
:= \frac{\nr{\{i \in \NN\, \mid  {E}_i(H_\omega^G) < {E} \}}}{ \nr{G}} .
\]
We will be in particular interested in the case where $G$ is a box $\Lambda_L=[-L,L]^d$. 
For a finite range hopping operator $H_0$, a box $\Lambda_L$ and a random configuration $\omega\in \Omega$ we use for brevity sake the following notation:
$H_\omega^L:=H_\omega^{\Lambda_L}$ and $N_\omega^L({E}):=N(H_\omega^L,{E})$.

Now we introduce the notion of $H_0$-connectedness induced by a finite hopping range operator $H_0$.
Two vertices $k,j \in \ZZ^d$ are $H_0$\emph{-nearest-neighbours} if $H_0(k,j)\neq 0$.
In this case we write $H_0\text{-}\dist(k,j)=1$. 
A $H_0$\emph{-path} (of length $n$) in $G\subset \ZZ^d$ is a
sequence of vertices $k_0, k_1, \dots, k_n\in G$ such that  $(k_0, k_1), \dots (k_{n-1},k_n)$ are pairs of $H_0$-nearest neighbours.
This induces the notion of $H_0$\emph{-path connected components}. The length of the shortest $H_0$-path joining $k$ and $j$ is denoted by $H_0\text{-}\dist(k,j)$.
If we write simply $\dist(k,j)$ we mean the distance function associated to the adjacency operator. 
For fixed $H_0$ and $\omega\in \Omega$ we denote by $X^\infty(\omega)$ the union of the infinite $H_0$-components of $X(\omega)$, and $\Lambda_L \cap X^\infty(\omega)$ by $\Lambda_L^\infty(\omega)$.
The restriction of a random Hamiltonian $H_\omega$ to $\Lambda_L^\infty(\omega)$ and the associated finite volume IDS are denoted by $H_\omega^{\infty,L}$ and  $N_\omega^{\infty,L}$, respectively. Similarly, $H_\omega^\infty$ is the restriction of $H_\omega$ to $X^\infty(\omega)$.
\smallskip

Denote by $\sigma_{disc}, \sigma_{ess}, \sigma_{ac}, \sigma_{sc}, \sigma_{pp}$
the discrete, essential, absolutely continuous, singular continuous, and pure point part of the spectrum, and by  $\sigma_{fin}$ the set of eigenvalues which posses an eigenfunction with finite support. Denote by $P_\omega(I)$ the spectral projection onto on interval $I$ associated to the operator $H_\omega$.

\begin{thm}
\label{t-exIDS}
There exists an $\Omega' \subset \Omega$ of full measure and subsets of the
real numbers $\Sigma$ and $ \Sigma_\bullet$, where $\bullet \in\{disc, ess, ac, sc, pp, fin\}$,
such that for all $\omega\in \Omega'$
\[
\sigma(H_\omega)=\Sigma \quad \text{ and } \quad \sigma_\bullet (H_\omega)= \Sigma_\bullet
\]
for any $\bullet = disc, ess, ac, sc, pp, fin$. Moreover, $\Sigma_{disc}=\emptyset$.
There exist a distribution function $N$ called \emph{integrated density of states} such 
that for all $\omega \in \Omega'$
\be
\label{t-d-IDS}
\lim_{L\to \infty} N_\omega^L (E) = N(E)
\ee
at all continuity points of $N$. The following \emph{trace formula} holds for the
IDS
\be
\label{e-traceformula}
N(E) = \EE \left \{ \la \delta_0, P_\omega (]-\infty, E[) \delta_0 \ra   \right \}
=  |\Lambda_L|^{-1} \EE \left \{ \Tr [\chi_{\Lambda_L} P_\omega (]-\infty, E[)  ]\right \} \, \, \forall \, L\in \NN .
\ee
The almost-sure spectrum $\Sigma$ coincides with the set of points of increase of the IDS
\be
\label{e-suppnu}
\Sigma = \{ {E}\in \RR \mid  N({E}+\epsilon) > N({E} -\epsilon) \text{ for all $\epsilon >0$}\} .
\ee
Analogous statements hold for $H_\omega^\infty$. The corresponding quantities are denoted by $N_\omega^{\infty,L},N^\infty, \Sigma^\infty$ and $ \Sigma_\bullet^\infty$, where $\bullet = disc, ess, ac, sc, pp, fin$.
\end{thm}
This is a special case of Theorem \ref{t-GexIDS}.
Similar results for the Dirichlet and Neumann Laplacians on the active clusters on $\ZZ^d$ corresponding to 
bond percolation have been obtained in \cite{KirschM-04}.
\medskip

Note that the trace formula \eqref{e-traceformula} holds for any $L\in \NN$.
This feature is very useful when studying properties which depend on a finite part of the configuration $\omega$, but if the size of this finite part is not know a priori.

The definition of $N^\infty$ is in some sense unsatisfactory, since 
$N_\omega^{\infty,L}$ depends on events which happen infinitely far away from the box $\Lambda=\Lambda_L$.
However, it can be shown that $N^\infty$ can be defined by an approximating sequence with better properties. 
Denote by $\partial_l^i \Lambda =\{k\in \Lambda\mid \dist(k, \Lambda^c) \le l\}$ the inner $l$-boundary of $\Lambda$ and similarly by $\partial_l^o \Lambda$ its outer $l$-boundary. Here $\Lambda^c$ denotes the complement of the set $\Lambda$.
Let $\Lambda^{con}(\omega)$ be the set of vertices in $\Lambda(\omega)$ which are connected by a $H_0$-path in $X(\omega)$ to $\partial_R^o \Lambda$. Now $\Lambda^{con}(\omega)$ depends only on the random variables with index in $\Lambda$ and its outer $R$-boundary. Denote by $H_\omega^{con,L}$ the restriction of $H_\omega$ to $\Lambda_L^{con}(\omega)$ and by $N_\omega^{con,L}({E})$ the corresponding normalised eigenvalue counting function.

\begin{prp}
\label{p-Ncon}
For almost all $\omega$, $\lim_{L\to \infty} N_\omega^{con,L} (E) = N^\infty(E)$ holds at all continuity points $E$ of $N^\infty$.
\end{prp}

\begin{rem}[Maximum of $N^\infty$] 
The maximal value of the IDS $N^\infty$ is an information which can be obtained using the same ideas as for the proof of Proposition \ref{p-Ncon}.
For the adjacency operator on $X(\omega)$, we have $\lim_{E\to \infty}N(E)= \EE\{\Tr[\chi_{\{0\} \cap X(\omega)}]\}=p$.
For the operator on $X^\infty(\omega)$
\[
\lim_{E\to \infty} N^\infty(E)= \EE\{\Tr[\chi_{\{0\} \cap X^\infty(\omega)}]\}=G(\infty):= \text{ density of the infinite cluster. }
\]
In the case of $N^\infty$ this answers a question posed in the Remark on p.~L1175 in \cite{ChayesCFST-86}.
\end{rem}
\medskip

Let $\nu, \nu^\infty$ be the measures associated with the distribution functions $N$ and $N^\infty$, respectively. Then equation \eqref{e-suppnu} can be stated as $\supp \nu=\Sigma$, respectively $\supp \nu^\infty=\Sigma^\infty$.  
In a similar way the supports  $\supp\nu_{pp}$ and $\supp\nu_{pp}^\infty$  of the pure point part of $\nu$, respectively $\nu^\infty$, can be characterised.  Set 
\begin{align}
\label{e-tildeSigma}
\tilde\Sigma := & \{ E \in \RR \mid \exists \text{ finite } G\subset \ZZ^d \text{ and } f \in \ell^2(G)
\text{ such that } H^G f=Ef\} .
\end{align}

\begin{thm}
\label{t-SigmaT} 
\begin{enumerate}[ \rm(i)]
\item
$ \Sigma_{fin}= \supp \nu_{pp}$. 
\item
If $q_0$ is a non-trivial random variable which takes only the values $0$ and $\infty$, 
we have $ \Sigma_{fin}=\tilde  \Sigma$.  
\item
If an infinite $H_0$-cluster exists almost surely we have $ \Sigma_{fin}^\infty  = \supp \nu_{pp}^\infty$. 
\item 
\label{i-adjec-op}
If, moreover, $H_0$ is the adjacency operator, then $\Sigma_{fin}^\infty  = \Sigma_{fin}$. 
\end{enumerate}
\end{thm}
\kom{What I want to prove: Call $\omega$ and atom of $\PP$ if $\PP(\{\omega\})>0$ and set 
\begin{align} 
\tilde\Sigma := \bigcup_{G \subset \, \ZZ^d \text{ finite} \atop \text{ appears as cluster with positive $\PP$-probability}} \quad \bigcup_{\omega \text{ atom of } \PP} \sigma(H_\omega^G)
\end{align}
Note that for independent random variables $q_k, k\in \ZZ^d$ 
\begin{align} 
\tilde\Sigma = \bigcup_{G \subset \, \ZZ^d \text{ finite}} \quad 
\bigcup_{\omega  \ : \ q_k(\omega) \in \, \supp \mu_{pp} \, \forall \, k \in G} \sigma(H_\omega^G)
\end{align}
If $q_0$ is Bernoulli distributed with $\PP (q_0(\omega)=0)=p$ and $\PP (q_0(\omega)=\infty)=1-p$, then 
$\tilde\Sigma = \bigcup_{G \subset \ZZ^d \text{ finite }} \sigma(H_0^G)$.  
\\
THEOREM: If $\mu(\{\infty\})> 0$, then $\Sigma_{fin}= \tilde\Sigma$, otherwise $\Sigma_{fin}= \emptyset$. 
\\
Probably the same statement holds if we drop the independence assumption.
}

Statement \rm(i) of Theorem \ref{t-SigmaT} is by no means restricted to percolation Hamiltonians.
Its application to more general graph Hamiltonians will be discussed elsewhere. 
Assertion \eqref{i-adjec-op} of Theorem \ref{t-SigmaT} is stated for the adjacency operator only. It seems that it can be extended to quite general finite hopping range Hamiltonians, which are invariant under an axial symmetry.

The characterisation (ii) of $\Sigma_{fin}$ by $\tilde \Sigma$ provides us with additional information.
Since the approximations $H_\omega^L$ converge in strong resolvent sense to $H_\omega$, 
the set $\tilde \Sigma$ is dense in the almost sure spectrum of $H_\omega$. If $H_0$ is the adjacency operator, $\tilde \Sigma$ is contained in the ring of algebraic integers. 
\medskip

Theorem \ref{t-exIDS} and Proposition \ref{p-Ncon} assert only convergence at the continuity points of the IDS. 
On the set of discontinuities of $N$ the convergence may not hold. This set can be quite large, as is seen in the case of the adjacency operator.
From the following Theorem \ref{t-logHoelder} we will see that the convergence of the finite volume approximations sometimes holds even at the discontinuities.  

\begin{thm}
\label{t-logHoelder}
Let $E$ be an algebraic number and $H_0$ a finite hopping range operator with integer coefficients. Assume that there is an $n\in \NN$ such that $q_0$ takes values in $\{0, \dots, n\} \cup \{\infty\}$.
 Then there exists a constant $C_E$ such that for all $\epsilon \in ]0,1[$, $L\in \NN$ and $\omega \in \Omega$:
\[
N_\omega^L(E+\epsilon) - N_\omega^L(E) \le \frac{C_E }{\log(1/\epsilon)} .
\]
The same statement applies to the restriction $H_\omega^\infty$ to the infinite active cluster $X^\infty(\omega)$.
\end{thm}
More generally, Theorem \ref{t-logHoelder} still holds, if one merely assumes that $H_0$ and $q$ take values in a finite subset of the integers of an algebraic number field, see \cite[Sec.~9]{Farber-98}. 
Such estimates have been used to analyse the IDS of (non-random) Harper operators on graphs and Laplacians on simplicial complexes in \cite{MathaiY-02,DodziukMY}. These papers establish right log-H\"older continuity of the IDS at algebraic integers and global convergence for the models considered there.
\medskip

In the case that $H_0$ is the adjacency operator, all discontinuities of the IDS are algebraic integers, and so we can use Theorem \ref{t-logHoelder} to derive the following 

\begin{cor}
\label{c-GlobConv}
If $H_0$ is the adjacency operator, the convergence $ \lim_{L\to \infty} N_\omega^L (E) = N(E)$ 
holds for all $E\in \RR$. Moreover, the IDS is right log-H\"older continuous at algebraic integers $E$
\[
N(E+\epsilon) - N(E) \le \frac{C_E }{\log(1/\epsilon)}
\]
where $\epsilon$ and $C_E$ are as in Theorem \ref{t-logHoelder}.
\end{cor}
\medskip

In \cite{DelyonS-84} Delyon and Souillard showed that the IDS of the Anderson model on the lattice is continuous, for any distribution of the coupling constants $q_k$. For this result, the coupling constants even do not need to be independent, but merely an ergodic array of random variables. In \cite{DelyonS-84} it is furthermore observed that this result breaks down for the quantum percolation model. We address now the question, how their result can be adapted to mixed Anderson-percolation Hamiltonians we are considering. 

\begin{thm}
\label{t-DelyonS}
Assume that $\mu=\mu_c +(1-p)\delta_\infty$, i.e.~$\mu$ has no atoms at finite values. Then the IDS of $H_\omega$ is continuous.
\end{thm}

Finally, we derive a Wegner estimate for finite truncations of the Hamiltonian $H_\omega$. It implies the Lipschitz continuity of the IDS and a  bound on its derivative $\frac{dN(E)}{dE}$, the \emph{density of states}. Wegner estimates \cite{Wegner-81}
play a crucial role in the study of the dense pure point spectrum of random Hamiltonians. Overviews dealing with this topic include \cite{CarmonaL-90,PasturF-92,Stollmann-01,Veselic-04a}, see also the references therein.

\begin{thm}
\label{t-Wegner}
Assume $\sigma(H) \subset [s_-,s_+]$ and that for $a,b\in \RR$ the measure $\mu$ is absolutely continuous on the interval $]a+s_-, b+s_+[$, i.e.~$\mu|_{]a+s_-, b+s_+[}(dx)=f(x)dx$, and that $f\in L^\infty$. Then, for every interval $I$ with  $\dist(I,]a,b[^c)\ge \delta>0$ we have
\be
\label{e-WE}
\EE \{\Tr \chi_I(H_\omega^L) \} \le C \, |I| \, L^d
\ee
where $C= 2^{d+2}\, \l(\frac{b-a+s_+-s_-+1}{\delta}\r)^2   \frac{\|f\|_\infty }{\mu (]a+s_-, b+s_+[)} $.
\end{thm}
It follows that the constant $C$ in \eqref{e-WE} is an upper bound on the density of states:
\[
\frac{dN(E)}{dE} \le C \text{ for all }  E \in ]a,b[
\]
The theorem applies in particular to distributions $\mu$ of the form $p\, f(x) dx+ (1-p)\, \delta_\infty(x)$.
If a bounded density exists globally for $\mu$, we recover global Lipschitz continuity of the IDS and a global bound on the density of states, cf.~\cite{Wegner-81}.

In \cite{KirschV-02a} a similar result has been proven by Kirsch and the present author for random  Schr\"odinger operators on $L^2(\RR^d)$ near the bottom of the spectrum. 
To separate the singular component of the distribution $\mu$ one has to use some ideas beyond the usual Wegner estimates. Similar ideas to those used in \cite{KirschV-02a} and the present paper can be found in \cite{Jeske-92}, respectively \cite[\S~3.2]{CombesHKN-02}.
\medskip

Both Theorems \ref{t-Wegner} and \ref{t-DelyonS}, together with their proofs, apply to the restriction  $H_\omega^\infty$ of the random Hamiltonian to the infinite cluster, too. 

\section{Generalisations: Amenable graphs and correlated potentials}
\label{s-general}
Most of the results stated in the previous section extends to more general models. In particular one can replace the lattice $\ZZ^d$ by a more general graph and relax the iid condition on the stochastic process $q_k(\omega)$.
In this section we explain this more general setting and state which of the theorems of \S~\ref{s-results}
hold in this situation. The proofs in Sections \ref{s-LocDef} to \ref{s-Wegner} are given for the general models.
More details about the model discussed in the sequel can be found in \cite{Veselic-QP}.
\bigskip

Let $V$ be the set of vertices and $E$ the set of edges of a graph $X$. 
Let $\dist\colon V\times V \to \NN \cup \{0\}$ be the distance function $V$ assigning to each pair of vertices the length of the shortest path between them. In the sequel we will by abuse of language identify the graph $X$ with its set of vertices $V$. Note that two vertices are connected by an edge if and only if their distance $\dist$ is equal to one. Thus the information contained in the set $E$ may be replaced by the information contained in the function $\dist$. We will be considering sub-graphs $G$ of $X$. The distance function on $G$ will be simply the restriction of the distance function on $X$, i.e.~we will only consider \emph{induced} sub-graphs of $X$.
Our situation is so simple because we are only considering site-percolation. Bond-percolation gives rise to 
general sub-graphs of $X$ which need not be induced.

Let $\Gamma$ be a group of graph-automorphism acting on the graph $X$.
It induces a projection map $\proj \colon X \to X / \Gamma$. We assume
that the quotient is a finite graph. This implies in particular that the
degree of the vertices in $X$ is uniformly bounded. We denote the smallest upper bound
by $\deg_+$.
Chose a vertex $[k]\in X/ \Gamma$ and a representative $k\in [k]
\subset X$. Starting from $k$, lift pathwise the vertices and edges of $X/ \Gamma$ to
obtain a connected set of vertices and edges $\tilde\cF\subset X$, such that $\proj|_{\tilde\cF}
\colon \tilde\cF \to X/ \Gamma$ is a bijective map. The set $\cF:= \tilde\cF \cup \{k \in X \mid k \text{ is an endpoint of an edge in } \cF \}$ is a graph, which we call \emph{fundamental domain}. Note that $\proj|_\cF
\colon \cF \to X/ \Gamma$ is a graph-map, which is bijective on the set of edges, but not on the set of vertices.
\medskip

We construct a probability space $(\Omega, \cA, \PP)$ associated
to percolation on $X$. Let $\Omega= \times_{k\in X} [0, \infty]$
be equipped with the $\sigma$-algebra $\cA$ generated by finite dimensional cylinders sets.
Denote by $\PP$ a probability measure on $\Omega$ and assume that the measurable shift transformations
\[
\tau_\gamma\colon \Omega \to \Omega, \quad 
(\tau_\gamma \omega)_k = \omega_{\gamma^{-1}k}
\]
are measure preserving. Moreover, let the family $\tau_\gamma , \gamma \in \Gamma$
act ergodically on $\Omega$.
By the definition of $\tau_\gamma , \gamma \in \Gamma$
the stochastic field $q\colon \Omega\times V \to [0,\infty]$
given by $q(\omega,k)=q_k(\omega)=\omega_k, k \in V$ is \emph{stationary} or \emph{equivariant},
i.e.{} $q(\tau_\gamma\omega,k)=q(\omega,\gamma^{-1}k) $.
The mathematical expectation associated to the probability $\PP$ will be denoted by $\EE$.

By the assumptions on  $X$ and $\cF$, the group $\Gamma$ is discrete and finitely generated.
We assume that $\Gamma$ is \emph{amenable}.  This means that there exists a 
sequence $\{I_L\}_L$ of finite, non-empty subsets of
$\Gamma$ such that for any finite $K\subset \Gamma$ and $\epsilon >0$
 \be
 \label{e-FS}
 \vert I_L \Delta K I_L \vert  \le \epsilon \,{\vert I_L\vert}
 \ee 
for all $ L$ large enough. A sequence $\{I_L\}_L$ with this property is called \emph{{F\o lner} sequence}, cf.~e.g.~\cite{AdachiS-1993}.  Now for finitely generated amenable groups there exists  a {F\o lner} sequence of subsets, 
which is increasing and exhausts $\Gamma$, cf.~Theorem 4 in \cite{Adachi-1993}. 
From \cite{Lindenstrauss-01} we infer that each {F\o lner} sequence has an tempered
subsequence. A \emph{tempered} {F\o lner} sequence is a sequence
which satisfies in addition to \eqref{e-FS} the growth condition
\[
\text{ there exists $C < \infty$ such that for all } L \in \NN \ : \
\nr{I_{L} I_{L-1}^{-1}} \le C \nr{I_{L}}
\]
where $I_L^{-1}:= \{\gamma |\gamma^{-1}\in I_L \}$. 
To each increasing, tempered {F\o lner} sequence associate an
\emph{admissible exhaustion} $\{\Lambda_L\}_L$ of $X$ given by
\be
\label{e-LambdaL}
\Lambda_L := \bigcup_{\gamma \in I_L^{-1}} \gamma \cF \subset X .
\ee

A (linear) finite hopping range operator $H_0\colon \ell^2(\Gamma)\to \ell^2(\Gamma)$ is defined by the properties 
\begin{enumerate}[\rm(i)]
\item $H_0(k,j) = H_0(j,k) \in \RR$,
\item $H_0(\gamma k,\gamma j) = H_0(k,j)$ for all  $\gamma  \in \Gamma$ and
\item $H_0(k,j)=0$ if $\dist(k,j) \ge R$, for some $ R\in \NN$. 
\end{enumerate} 
Assume without loss of generality $|H_0(k,j)| \le 1$
for all matrix elements. It follows that the $\ell^2$-operator-norm of $H_0$ is bounded
by $M:=2 \deg_+^{R}$. Since $H_0$ is symmetric, it is a selfadjoint
operator. In particular, the spectrum of $H_0$ and all its restrictions to sub-graphs of $X$ is contained in $[-M,M]$. 

\kom{The following table illustrates the relation between the general situation considered in the present section and the model of Section \ref{s-results}.  \\[.2em]
\begin{tabular}{|l| |c|c| } 
    \hline     
    &  general situation & specific model 
    \\ 
    &  Section \ref{s-general} & Section \ref{s-results} 
    \\    \hline\hline 
    space & $X$ graph with amenable $\Gamma$-action& $\ZZ^d$ as graph  
    \\ \hline
    exhaustion & constructed from & sequence of cubes
    \\ 
    $\Lambda_L, L \in \NN$ & F\o lner sequence $I_L, L\in \NN$ & $\Lambda_L=[-L,L]^d$
    \\ \hline
    Hamiltonian & & 
    \\     \hline
    invariant part & $H_0$ real symmetric, $\Gamma$-invariant  & $A$ adjacency 
    \\ & finite hopping range operator & operator 
    \\ \hline
    random part &      $q_k \in [0, \infty], k\in X$ &  $q_k \in \{0, \infty\}, k\in \ZZ^d$
    \\      \hline 
    correlation & $\Gamma$-ergodic stochastic process & i.i.d. process 
    \\ \hline
\end{tabular} \smallskip }

The sub-graphs $X(\omega)$, $X^\infty(\omega)$, $\Lambda_L$, $\Lambda_L(\omega)$, $\Lambda_L^\infty(\omega)$, $\Lambda_L^{con}(\omega)$ and the operators $H_\omega$, $H_\omega^L$, $H_\omega^{\infty,L}$, $H_\omega^{con,L} $ are defined as in Section \ref{s-results}, except that $\ZZ^d$ is replaced by the graph $X$ and $\Lambda_L$ is defined by \eqref{e-LambdaL}.

In the remainder of this section we present generalisations of the results of \S~\ref{s-results} to the more general setting introduced above.
\smallskip

\begin{thm}
\label{t-GexIDS}
There exists an $\Omega' \subset \Omega$ of full measure and subsets of the
real numbers $\Sigma$ and $ \Sigma_\bullet$, where $\bullet \in\{disc, ess, ac, sc, pp, fin\}$,
such that for all $\omega\in \Omega'$
\[
\sigma(H_\omega)=\Sigma \quad \text{ and } \quad \sigma_\bullet (H_\omega)= \Sigma_\bullet
\]
for any $\bullet = disc, ess, ac, sc, pp, fin$. Moreover,  $\Sigma_{disc}=\emptyset$ for infinite $\Gamma$.
There exist a distribution function $N$ called \emph{integrated density of states} such 
that for all $\omega \in \Omega'$
\be
\label{t-d-GIDS}
\lim_{L\to \infty} N_\omega^L (E) = N(E)
\ee
at all continuity points of $N$. The following \emph{trace formula} holds for the
IDS
\be
\label{e-Gtraceformula} 
N(E) = |\cF|^{-1} \EE \left \{ \Tr[ \chi_\cF P_\omega (]-\infty, E[) ]   \right \}
=  |\Lambda_L|^{-1} \EE \left \{ \Tr [\chi_{\Lambda_L} P_\omega (]-\infty, E[)  ]\right \} \, \, \forall \, L\in \NN .
\ee
The almost-sure spectrum $\Sigma$ coincides with the set of points of increase of the IDS
\be
\label{e-Gsuppnu}
\Sigma = \{ {E}\in \RR \mid  N({E}+\epsilon) > N({E} -\epsilon) \text{ for all $\epsilon >0$}\}
\ee
Analogous statements hold for the quantities $N_\omega^{\infty,L},N^\infty, \Sigma^\infty$ and $ \Sigma_\bullet^\infty$, where $\bullet = disc, ess, ac, sc, pp, fin$, which are associated to $H_\omega^\infty$.
\end{thm}
The proof of the theorem can be inferred from \cite{Veselic-QP}. It is based on the same argument for random lattice Hamiltonians. The difference to this simpler case is that the group $\Gamma$ is not abelian, but merely amenable.
To overcome this difficulty some techniques from \cite{PeyerimhoffV-02,LenzPV-04} are used. 
Equality \eqref{e-Gtraceformula} is related to the fact that \emph{all} means on $\ell^\infty(\Gamma)$ take the same value on the function
\[
X \ni k \mapsto \EE \{\Tr[\delta_k P_\omega(]-\infty ,E[)] \} \in \RR
\]

\begin{prp}
\label{p-GNcon}
For almost all $\omega$, $\lim_{L\to \infty} N_\omega^{con,L} (E) = N^\infty(E)$ holds at all continuity points $E$ of $N^\infty$.
\end{prp}

The set $\tilde\Sigma$ is defined in the same way as in \eqref{e-tildeSigma}, except that $\ZZ^d$ is now replaced by $X$.

\begin{thm}
\label{t-GSigmaT}
\begin{enumerate}[ \rm(i)]
\item
$ \Sigma_{fin}= \supp \nu_{pp}$. 
\item
If $q_k, k\in X$ take only the values $0$ and $\infty$, we have $ \Sigma_{fin} \subset\tilde  \Sigma$.  
If, moreover, the $q_k, k \in X $ are independent and non-trivial random variables, 
we have $ \Sigma_{fin}=\tilde  \Sigma$. 
\item
If an infinite $H_0$-cluster exists almost surely we have $ \Sigma_{fin}^\infty  = \supp \nu_{pp}^\infty$. 

\end{enumerate}
\end{thm} 

The proof of the equality $\Sigma_{fin}^\infty  = \Sigma_{fin}$ for $X=\ZZ^d$, $H_0$ the adjacency operator, and $q_k, k \in \ZZ^d$ iid, does not extend directly to more general situations. Certainly it is necessary that the graph exhibits sufficiently many axes of symmetry, as is for example the case for the triangular or the honeycomb lattice. It is also not immediately clear how to extend the result to correlated $q_k$. One should expect the result to hold, if the range of the Hamiltonian is larger than the correlation length of the process $q_k$.
\smallskip

\begin{thm}
\label{t-GlogHoelder}
Let $E$ be an algebraic number and $H_0$ a finite hopping range operator with integer coefficients. Assume that there is an $n\in \NN$ such that $q_0$ takes values in $\{0, \dots, n\} \cup \{\infty\}$.
 Then there exists a constant $C_E$ such that for all $\epsilon \in ]0,1[$, $L\in \NN$ and $\omega \in \Omega$:
\[
N_\omega^L(E+\epsilon) - N_\omega^L(E) \le \frac{C_E }{\log(1/\epsilon)} .
\]
The same statement applies to the restriction to the infinite active cluster $X^\infty(\omega)$.
\end{thm}
\begin{cor} \label{c-GGlobConv}
Corollary \ref{c-GlobConv} applies verbatim the the present, more general situation.
\end{cor}

For deterministic graphs, i.e.~in the case that the probability space degenerates to a point,  such results have been obtained in \cite{DodziukMY}. 
In this case our proof of Corollary \ref{c-GGlobConv} seems to streamline some arguments from the $L^2$-invariants literature, e.g.~\cite{DodziukMY,MathaiY-02}. For a different approach to prove the convergence of the IDS at all energies for periodic Hamiltonians see \cite{MathaiSY-03}.
\smallskip

Theorems \ref{t-DelyonS} and \ref{t-Wegner} are special cases of the following results:

\begin{thm}
\label{t-GDelyonS}
Assume that the random variables $q_k, k \in X$ are independent and $\mu=\mu_c +(1-p)\delta_\infty$, i.e.~$\mu$ has no atoms at finite values. Then the IDS of $H_\omega$ is continuous.
\end{thm}

\begin{thm}
\label{t-GWegner}
Assume that the random variables $q_k, k \in X$ are independent, $\sigma(H_0) \subset [s_-,s_+]$ and that for $a,b\in \RR$ the measure $\mu$ is absolutely continuous on the interval $]a+s_-, b+s_+[$, i.e.~$\mu|_{]a+s_-, b+s_+[}(dx)=f(x)dx$, and that $f\in L^\infty$. Then, for every interval $I$ with  $\dist(I,]a,b[^c)\ge \delta>0$ we have
\be
\label{e-GWE}
\EE \{\Tr \chi_I(H_\omega^L) \} \le C \, |I| \, |\Lambda_L|
\ee
where $C= 2^{d+2}\, \l(\frac{b-a+s_+-s_-+1}{\delta}\r)^2   \frac{\|f\|_\infty }{\mu (]a+s_-, b+s_+[)} $.
\end{thm}

\section{Local definition of the IDS}
\label{s-LocDef}
In this section we prove Proposition \ref{p-GNcon}, i.e.~we show that the IDS on
the infinite cluster can be defined in two equivalent ways. The
first one is defined by considering finite volume restrictions of
the infinite active cluster; the second by using only local
information to define the finite volume operator.

On the way we get insights which are by themselves
interesting.  We prove that finite clusters in $X(\omega)$ have a
well defined density. Moreover, the density of clusters of size
$n$ tends to zero  as $n$ tends to infinity.

For $n\in \NN$ set
\[
X^n(\omega):= \{k \text{ is contained in a finite active cluster of size $\ge n$ } \} .
\]
For both $n \in \NN$ and $n = \infty$ denote
$g(n,\omega,L):=\frac{|X^n(\omega)\cap\Lambda_L| }{|\Lambda_L|}$.

\begin{lem}
For $n \in \NN \cup \{\infty\}$ there exists $G(n)\in [0,1]$ such that 
\[
\lim_{L \to \infty} g(n,\omega,L)= G(n) \quad \text{ for almost all $\omega$. } 
\]
\end{lem}
\begin{proof}
\begin{align*}
g(n,\omega,L)
&= |\Lambda_L|^{-1}  \sum_{k\in X} \chi_{X^n(\omega)}(k) \chi_{\Lambda_L}(k)
&= |I_L|^{-1} \sum_{\gamma\in I_L^{-1}} |\cF|^{-1}   \sum_{k\in X} \chi_{X^n(\omega)}(k) \chi_{\gamma \cF}(k) .
\end{align*}
Since $\chi_{\gamma \cF}(k)=\chi_{\cF}(\gamma^{-1}k)$,
we have
\[
\sum_{k\in X} \chi_{X^n(\omega)}(k) \chi_{\gamma \cF}(k) =
\sum_{j\in X} \chi_{X^n(\omega)}(\gamma j) \chi_{\cF}(j) .
\]
Now $\gamma X(\omega)= X(\tau_\gamma \omega)$ implies $\gamma X^n(\omega) = X^n(\tau_\gamma \omega)$, and this in turn $\chi_{X^n(\omega)}(\gamma j) = \chi_{\gamma^{-1} X^n(\omega)}(j)=\chi_{X^n(\tau_\gamma^{-1} \omega)}(j)$.
With
\[
g(n, \omega):= |\cF|^{-1} \sum_{j\in \cF} \chi_{X^n(\omega)}(j)= |\cF|^{-1} \Tr(\chi_{\cF} \chi_{X^n(\omega)})
\]
this implies  
$ \displaystyle |\cF|^{-1} \sum_{j\in X} \chi_{X^n(\omega)}(j) \chi_{\gamma \cF}(j) = g(n, \tau_\gamma^{-1} \omega)$.
Now $0\le g(n,\omega) \le 1$ for all $n$ and $\omega$, thus the function is certainly in $L^1(\Omega)$. 
Thus we may apply Lindenstrauss' ergodic theorem \cite{Lindenstrauss-01} and conclude for almost every $\omega \in \Omega$, 
as well as in $L^1$ sense the following convergence:
\[
g(n,\omega,L) 
= |I_L|^{-1} \sum_{\gamma\in I_L^{-1}} g(n,\tau_\gamma^{-1}\omega) 
\to \EE\{ g(\cdot,n)\} 
= |\cF|^{-1} \EE\{ \Tr(\chi_{\cF} \chi_{X^n(\cdot)})\}
=: G(n)
\]
 for $L\to \infty$.
\end{proof}

This proves that clusters of size $n$ have a well defined density. Indeed, since $X^n\supset X^{n+1}$ the density of clusters of size $n$ is just $G(n)-G(n+1)$. 
Note that due to the the way how $X^n$ is defined we do not have $G(n) \to G(\infty)$ unless the infinite cluster is empty almost surely.
Next we prove that the density tends to zero as $n$ grows
:\[
G(n) \to 0 \quad \text{ as } n \to \infty .
\]
For this aim, we will show that $g(\omega, n)= |\cF|^{-1} \Tr[\chi_{\cF} \, \chi_{X^n(\omega)} ] $ converges to zero pointwise.

Observe the monotonicity  $X^n(\omega) \supset X^{n+1}(\omega)$ and $\cap_n X^n(\omega)=\emptyset$.
In particular, for each $k\in X$ there exists a $N_k$ such that $ k \not \in X^n(\omega)$ for all $n\ge N_k $.
Choose $N=\max_{k\in \cF}N_k$, then $\cF \cap X^n(\omega) =\emptyset $ for all $n \ge N$.
\[
\Ra \Tr [ \chi_{\cF} \chi_{X^n(\omega)}]  =0 \quad \text{ for } n\ge N .
\]
Note that $N$ depends on $\omega$. Since $0\le g(\omega, n)\le 1$
and $\PP$ is finite, Lebesgues' dominated convergence theorem implies $\lim_{n\to \infty } \EE\{g(\cdot,n) \}=0$.
\bigskip

Now we turn to the proof of Proposition \ref{p-GNcon}. For any random Hamiltonian $B_\omega$ and $\Lambda \subset X$ set $\Lambda^\infty (\omega):= \Lambda \cap X^\infty (\omega)$,
\[
B_{\Lambda(\omega)}^\infty = \chi_{\Lambda^\infty(\omega)} B_\omega\chi_{\Lambda^\infty(\omega)}
\]
and
\[
B_{\Lambda(\omega)}^{con} = \chi_{\Lambda^{con}(\omega)} B\chi_{\Lambda^{con}(\omega)}
\]
where $\Lambda^{con}(\omega):= \{k\in \Lambda \mid \text{ exists path of active sites from $k$ to } \Lambda^c   \} $, as defined before.
The equality of $N^\infty $ and $N^{con}$ will follow from the following
\begin{lem}
For almost all $\omega\in \Omega$ we have
\[
\lim_{L \to \infty} \frac{\Tr [B_{\Lambda_L(\omega)}^{con} - B_{\Lambda_L(\omega)}^\infty ] }{|\Lambda_L|} = 0 .
\]
\end{lem}
\begin{proof}
We have 
\[
\Tr [B_{\Lambda_L(\omega)}^{con} - B_{\Lambda_L(\omega)}^\infty ]
 =
\sum_\bullet [B(k,k) ]
\]
where the bullet denotes summation over all active sites in $\Lambda_L$ which are not in $X^\infty(\omega)$, but are  connected by an active path to $\Lambda_L^c$. 
Thus 
\begin{align*}
\frac{\Tr [B_{\Lambda_L(\omega)}^{con} - B_{\Lambda_L(\omega)}^\infty ]}{|\Lambda_L|}
& \le
\|B\| \left ( g(h,\omega,L) + \frac{|\partial_h^i \Lambda_L| }{|\Lambda_L|}  \right )
\end{align*}
converges to $ \|B\| \ \EE\{g(h,\omega)\}=  \|B\| \ G(h)$ as $L\to \infty $. However, $G(h)$ goes to zero for 
$h \to \infty$, as we   saw before.
\end{proof}

The lemma can be applied to $B_\omega=H_\omega^m$. Thus it implies that  the difference of the moments of $N_\omega^{con, L}$ and $N_\omega^{\infty, L}$ converges to zero almost surely as $L\to \infty$. Thus the sequence $N_\omega^{con, L}$  converges also to the distribution function $N$, which proves Proposition \ref{p-GNcon}.

\section{Finitely supported eigenstates \& discontinuities of the IDS}
\label{s-finitestates}
In this section we prove Theorems \ref{t-SigmaT} and \ref{t-GSigmaT}. Along the way we establish a correspondence between finitely supported eigenstates, unique continuation properties,  and discontinuities of the IDS. 
\bigskip

\begin{dfn}
For a given graph $X$, a probability measure $\PP$ on the corresponding probability space $\Omega$, a finite hopping range operator (of range $R$), and a random Hamiltonian $H_\omega$, we say that the \emph{unique continuation property holds for $\{H_\omega\}_\omega$ at energy $E\in \RR$} if for any finite $\Lambda\subset X$, $\PP$-almost every $\omega$ and $f \in \ell^2(X(\omega))$
\[
f\equiv 0 \text{ on } \partial_{2R}^o \Lambda \text{ and } H_\omega f=Ef \text{ implies } f\equiv 0 \text{ on } \Lambda.
\]
The \emph{unique continuation property for $\{H_\omega^\infty\}_\omega$} is defined analogously.
\end{dfn}

\begin{prp}
\label{p-equiv}
Let $X,\Omega,\PP$ and $H_\omega$ be as above. The following properties are equivalent
\begin{enumerate}[\rm(i)]
\item
The IDS of $\{H_\omega\}_\omega$  is discontinuous at $E$.
\item
The unique continuation property does not hold for $\{H_\omega\}_\omega$ at $E$.
\item
$E\in \Sigma_{fin}$, where $\Sigma_{fin}$ is associated to $\{H_\omega\}_\omega$.
\end{enumerate}
\end{prp}

For the family $\{H_\omega^\infty\}_\omega$ the statement of Proposition \ref{p-equiv} holds analogously.

\begin{proof}
We prove first $\lnot (ii) \Rightarrow \lnot (i)$, i.e.~the unique continuation property at $E$ implies the continuity of $N$ at $E$. Let $\Lambda\subset X$  be finite and $f \in \ran \chi_\Lambda P_\omega(E)$, i.e.~ $f=\chi_\Lambda g, H_\omega g =Eg$.
If $g\equiv 0$ on $\partial_{2R}^o \Lambda$ then the unique continuation property implies $g\equiv0$ on $\Lambda$ and thus $f\equiv0$ on $X(\omega)$.
Thus it follows: $\dim \ran \chi_\Lambda P_\omega(E) \le \dim \ell^2(\partial_{2R}^o\Lambda)= |\partial_{2R}^o\Lambda| $.
\smallskip

Since the dimension of the range and the co-kernel coincide
\[
\Tr(\chi_\Lambda P_\omega(E)) = \Tr(\chi_\Lambda P_\omega(E)\chi_\Lambda ) \le \|\chi_\Lambda P_\omega(E)\chi_\Lambda \| \dim \ran \chi_\Lambda P_\omega(E)\chi_\Lambda \le |\partial_{2R}^o\Lambda| .
\]
This holds for any $\Lambda=\Lambda_L$ and almost every $\omega$. Thus
\[
|\cF|^{-1} \EE \{\Tr [\chi_{\cF}P_\omega(E)] \}
=
\inf_L |\Lambda_L|^{-1} \EE \{\Tr [\chi_{\Lambda_L}P_\omega(E)]\} =0
\]
since $\Lambda_L,L\in \NN$ is based on a F\o lner sequence.

We prove $(ii) \Rightarrow (iii)$.
Since the unique continuation property fails at $E$, there exists a finite $\Lambda \subset X$ and $\Omega' \subset \Omega$ of positive measure, such that for all $\omega \in \Omega'$ there is function $f_\omega\in \ell^2(V)$ with the
following properties:
\[
f_\omega \equiv 0 \text{ on } \partial_{2R}^o \Lambda \qquad f_\omega \not\equiv 0 \text{ on } \Lambda
\quad \text{and}\quad H_\omega f_\omega = E f_\omega .
\]
Now $g_\omega := \chi_\Lambda f_\omega$ is a finitely supported eigenfunction for $E$, thus $E\in \Sigma_{fin}$.

We prove $(iii) \Rightarrow (i)$.
Since $E\in \Sigma_{fin}$ there is $\Omega'\subset \Omega$ of full measure such that $E\in \sigma_{fin}(H_\omega)$ for all $\omega \in \Omega'$. Set $\Omega_{E,L}:= \{\omega \mid \exists f \in  \ell^2(X): \supp f \subset \Lambda_L \text{  and } H_\omega f = Ef  \}$.
Then
\[
\Omega' \subset \Omega_E :=\{ \omega \mid \exists \Lambda \subset X \text{ finite and } f \in \ell^2(X): \supp f \subset \Lambda, H_\omega f = Ef  \}   = \cup_{L \in \NN} \Omega_{E,L} .
\]
Since $\PP(\Omega_E)=1$, not all $\Omega_{E,L}$ can have measure $0$. Thus there is a $J\in \NN$ such that  
$\PP(\Omega_{E,J}) >0$.
By definition there exists for all $\omega \in \Omega_{E,J}$ a normalized $f\in \ell^2(X)$ with support contained in $\Lambda_J$ and $H_\omega f=Ef$.
We decompose the spectral projection $P_\omega(E) = |f\ra \la f| + \tilde P_\omega(E)$, $\tilde P_\omega(E) \ge 0$.
Consequently $\chi_{\Lambda_J} P_\omega(E) \chi_{\Lambda_J} \ge |f\ra \la f|$ ,
\[
\Tr[\chi_{\Lambda_J} P_\omega(E) ] \ge \Tr[|f\ra \la f| ] = \sum_{k\in \Lambda_J} |f(k)|^2=1 \text{ for } \omega \in \Omega_{E,J}
\]
and $ |\cF|^{-1}\EE\{\Tr[\chi_{\cF} P_\omega(E) ]  \} =
|\Lambda_J|^{-1}\EE\{\Tr[\chi_{\Lambda_J} P_\omega(E) ]  \} \ge |\Lambda_J|^{-1}\PP(\Omega_{E,J})>0     $.
\end{proof}
Similar arguments to the conclusion $\lnot (ii) \Rightarrow \lnot (i)$ have been used elsewhere, see e.g.~\cite{DelyonS-84,MathaiY-02,KlassertLS-03,DodziukMY}.

\begin{prp}
If $H_0$ is the adjacency operator on $\ZZ^d$ and the $q_j, j\in \ZZ^d$ are iid, we have $\Sigma_{fin}^\infty  = \Sigma_{fin}$. 
\end{prp}

\begin{proof}
The inclusion $\Sigma_{fin}^\infty  \subset \Sigma_{fin}$ is trivial. 

For $E\in \Sigma_{fin}$ there exists a finite set $S\subset \ZZ^d$ such that 
\[
\Omega_S :=\{\omega \mid \exists f_\omega \in \ell^2(X(\omega)), \supp f_\omega =S, H_\omega f_\omega=Ef_\omega \}
\]
has positive measure. Pick a choice 
\be
\label{e-OS}
\omega\mapsto f_\omega \text{ for all } \omega \in \Omega_S
\ee
 Set $a:= \min\{j_1\mid j\in S\}$ and let $\Lambda' \subset \ZZ^d$ be the minimal rectangular box which contains $S\cup \partial_2^o S\setminus \{j\mid j_1=a-2,a-1\}$. Define the reflection 
\[
R\colon \ZZ^d \to \ZZ^d, \quad R(j_1,\dots,j_d)=(2a-j_1,j_2\dots,j_d) .
\]
The set $\Lambda:= \Lambda' \cup R(\Lambda') \cup\{j\mid j_1=a-1, j \in \partial_1^0 \Lambda\}$ is a box.
Set $S^+:=S\cup\partial_1^oS \setminus \{j\mid j_1=a-1\}$ and
\begin{align*}
\Omega_R:  = \{\omega \in \Omega\mid \exists \omega' \in \Omega_S: \, 
& q_j = q_j' \, \forall \, j \in S^+
\\
& q_j = q_{R(j)}  \, \forall \, j \in R(S^+)
\\
& q_j < \infty \, \forall \, j \in \Lambda \setminus (S^+ \cup R(S^+)) 
\} .
\end{align*}
Since $\PP(\Omega_S) >0, \PP(q_0<\infty) >0$ and the random variables are i.i.d., it follows $\PP(\Omega_R) >0$. 
For each $\omega \in \Omega_R$ there exists an $\omega'\in \Omega_S$ with $\omega\upharpoonright_{S^+}=\omega'\upharpoonright_{S^+}$ and therefore a function $f_{\omega'}$ as in \eqref{e-OS}. We construct a new function 
\begin{equation*}
g_\omega(j):= \begin{cases}
              f_{\omega'}(j)  &\text{ for } j \in S^+\\
              f_{\omega'}(R(j)) &\text{ for } j \in R(S^+)\\
              0 &\text{ for all other } j \in X(\omega)
              \end{cases}
\end{equation*}
which satisfies $H_\omega g_\omega =Eg_\omega$. Note that $\Omega_R$ depends only on coordinates of $q$ with index  $j \in \Lambda$.

Pick a $k\in \partial_1^i\Lambda$ and consider the percolating regime where $\PP\{\Omega_k \} >0 $ for $\Omega_k=\{\omega\mid k \in X^\infty(\omega)\}$. If we pick an $\omega \in \Omega_k$ and replace the $j\in \Lambda$ coordinates with those of an element in $\Omega_R$, we still get a configuration in $\Omega_k$, since for elements 
in $\Omega_R$ all sites in $\partial_1^i\Lambda$ are active. By independence we conclude that $\supp g$ is a subset of $X^\infty(\omega)$ with positive probability.
\end{proof}
The mirror charge idea employed in the proof has been used previously in \cite{ChayesCFST-86}, and implicitly in a rudimentary form already in \cite{KirkpatrickE-72}.

\begin{prp}
\label{p-tS}
Let $H_0$ be finite hopping range Hamiltonian on a graph $X$ with amenable group $\Gamma$ action.
If the $q_k, k \in X$ are non-trivial, independent and take only the values $0$ and $\infty$, then $\Sigma_{fin} =\tilde\Sigma$.
\end{prp}

\begin{proof}
If $q_0 \in \{0,\infty\}$, then by definition of $\Sigma_{fin}$ we have $\Sigma_{fin}\subset \tilde \Sigma$. 
On the other hand by independence all finite $G\subset \ZZ^d$ occur in $X(\omega)$ with positive probability. Thus $\sigma(H_0^G)\subset \Sigma_{fin}$.
\end{proof}

\section{Log-H\"older continuity at algebraic numbers}
\label{s-lH}
In this section we prove Theorem \ref{t-GlogHoelder} and its Corollary \ref{c-GGlobConv}. The proofs rely on techniques of L\"uck \cite{Lueck-94c} and Farber \cite{Farber-98}. L\"uck studies in his paper the approximation of geometric $L^2$-invariants, in particular Betti numbers,  on covering spaces by their analogues on compact quotients. Farber works in a more general and abstract setting studying von Neumann categories. The techniques of these two papers have thereafter been used in different contexts, for instance for rational Harper operators on graphs \cite{MathaiY-02} or combinatorial Laplacians on covering spaces \cite{DodziukMY}.

The following is a formulation of a Lemma of L\"uck \cite{Lueck-94c}:
\begin{lem}
\label{l-Lueck}
Let $A\colon \CC^D\to \CC^D$ or $\colon \RR^D\to\RR^D$ be a hermitian, respectively symmetric, matrix and $p(t)=\det (t-A)$ its characteristic polynomial. Let $p(t)=t^kq(t)$ for a polynomial $q$ with $q(0)\neq 0$. Let $K=\max(1, \|A\|)$ and $0<C\le |q(0)|$. Denote by $\cN(A,E)$ the number of eigenvalues of $A$ (counting multiplicities) less than or equal to $E$. Then we have for all $\epsilon \in]0,1[$ 
\be
\label{e-logHoelder}
\cN(A,\epsilon) - \cN(A,0) \le \frac{\log 1/C+D\log K}{\log 1/\epsilon} .
\ee
\end{lem}
Note the different normalization of $\cN$ than in the case of finite volume restrictions of $H_\omega$.
The proof in \cite{Lueck-94c} applies to non-negative matrices $A$. Since in the following we will be dealing with merely symmetric matrices, we modify the proof to cover this case as well.

\begin{proof}
We enumerate the eigenvalues ${E}_i$ of $A$ in non-decreasing order counting multiplicities. Then there exist integers $p,r$ and $s$ such that 
\[
{E}_1 \le \dots {E}_p <0={E}_{p+1} = \dots = {E}_r < {E}_{r+1} \le \dots \le {E}_s \le \epsilon < {E}_{s+1} \le \dots  \le {E}_D .
\]
By definition $\cN(A,\epsilon) - \cN(A,0)=s-r$ and  $  \prod_{i=1}^p (t- {E}_i) \prod_{i=r+1}^D (t- {E}_i)=q(t)$, in particular
\[
\prod_{i=r+1}^s |{E}_i|= |q(0)| \, \prod_{i=1}^p |{E}_i|^{-1} \prod_{i=s+1}^D |{E}_i|^{-1} .
\] 
By definition of $s,C$ and $K$ we have
\[
\epsilon^{\cN(A,\epsilon) - \cN(A,0)}= \prod_{i=r+1}^s \epsilon \ge C \, K^{-D} .
\]
We take logarithms and obtain $\{\cN(A,\epsilon) - \cN(A,0) \}\log \epsilon \ge \log C - D\log K $. Since $\log \epsilon $ is negative, this implies
\[
\cN(A,\epsilon) - \cN(A,0) \le \frac{\log C-D\log K}{\log  \epsilon} .
\]
\end{proof}

If $A$ has integer coefficients, $|q(0)|$ is a non-vanishing integer, thus greater or equal to $1$.
One can apply the previous lemma to some energy $E$ which is not zero, if it is an algebraic number. For this the following facts will be useful:

Let $\alpha$ be an algebraic integer and $m_\alpha$ its monic minimal polynomial, i.e.~the irreducible polynomial with leading coefficient equal to one such that $m_\alpha(\alpha)=0$. The degree of $\alpha$ equals the degree of $m_\alpha$. The field $\QQ(\alpha)$ is an extension of $\QQ$ of degree $n$.
Enumerate all the roots $\alpha_1=\alpha, \alpha_2, \dots, \alpha_n$ of $m_\alpha$. Then there are $n$ 
distinct \emph{embeddings}  $\mathfrak{e}_j\colon \QQ(\alpha) \to \CC, j=1, \dots, n$ such that $\mathfrak{e}_j(\alpha)=\alpha_j$. The embeddings are homomorphisms of fields. 
The product of all roots of $m_\alpha$ can be written as $\prod_j \alpha_j =\prod_j \mathfrak{e}_j(\alpha)$. It is called the \emph{norm} of $\alpha$. Since it is the last coefficient of $m_\alpha$, it is an integer, and since $m_\alpha$ is irreducible, it does not vanish. Therefore
\be
\label{e-lowerbound}
|\mathfrak{e}_k(\alpha)| \ge \prod_{j\neq k} |\mathfrak{e}_j(\alpha)|^{-1} \ge \max_j\{\mathfrak{e}_j(\alpha)\}^{-n+1} .
\ee
Thus an upper bound on all $\mathfrak{e}_j(\alpha)$, implies a lower bound on all of them. This fact was first used in a similar context in \cite[Sec.~12.3]{Farber-98}.
\smallskip
 
Let $A$ be an $D\times D$ matrix with coefficients in $\ZZ$ and $E$ an algebraic number. Denote by $p(t)=\det(t-A)$ the characteristic polynomial of $A$. There is an integer $k\in \{0,\dots,D\}$ and a polynomial $q$ such that $p(t+E)=q(t)\, t^k$ and $q(0)\neq 0$. The value of $q(0)$ equals the coefficient of $t^k$ in the polynomial $p(t+E)$.
By expanding the polynomial one calculates (see e.g.~page 130 in \cite{MathaiY-02})
\[
q(0)= \sum_{j=0}^{D-k}  \l (\iznad{k+j}{k}\r ) c_{k+j} E^j
\]
where $c_r$ is the $r$-th coefficient of $p$, i.e.~the $r$-th symmetric polynomial of the roots of $p$. In particular $c_r \in \ZZ$.

There is an algebraic integer $\alpha$ and  $b\in \NN$ such that $E=\frac{\alpha}{b}$. (We may assume $\alpha\neq 0$, since we are preparing to apply Lemma \ref{l-Lueck}  to the operator $A-E$. For $\alpha=0$, $A-E=A$ 
and thus the Lemma may be applied directly.)
Thus $q(0)b^D$ is an algebraic integer. Since the embedding $\mathfrak{e}_l\colon \QQ(\alpha) \to \CC$ is an homomorphism of fields and thus leaves $\QQ$ invariant we have
\[
\mathfrak{e}_l(q(0)b^D) =  \sum_{j=0}^{D-k}  \l (\iznad{k+j}{k}\r ) c_{k+j} \, \mathfrak{e}_l(\alpha)^j \, b^{D-j} .
\]
To estimate the absolute value of this expression from above note the following:
\begin{enumerate}[\rm(i)]
\item
$\l(\iznad{k+j}{k}\r )\le 2D \, 4^D$, since $k,j \le D$. 
\item
$|\Tr(A^r)| \le D \|A\|^r$.
\item
Using Lemma B in \S~12.3 of \cite{Farber-98} we obtain
$|c_r| \le \l(\iznad{D}{r}\r)  \|A\|^r$.
\item
$\cR:=\max \{|\mathfrak{e}_l(\alpha)| \mid 1\le l \le n\} \ge |\mathfrak{e}_l(\alpha)| $ for all $1\le l \le n$.
\end{enumerate}
Thus $ |\mathfrak{e}_l(q(0)b^D) | \le 4 D^3 (8 \|A\| \cR b)^D$ and \eqref{e-lowerbound} implies 
\be
\label{e-q0}
| q(0) | \ge b^{-D}\{4 D^3 (8 \|A\| \cR b)^D\}^{-n+1} .
\ee

\bigskip
If we apply Lemma \ref{l-Lueck} to $A-E$ where $A$ and $E$ are as above, we obtain
\begin{align*}
\frac{\cN(A,E+\epsilon) - \cN(A,E)}{D} = \frac{\cN(A-E,\epsilon)- \cN(A-E,0) }{D} 
\le \frac{\log 1/C}{D\log 1/\epsilon}    +\frac{\log K}{\log 1/\epsilon}
\end{align*}
where $1/C$ can be chosen as $b^{D}\{4 D^3 (8 \|A\| \cR b)^D\}^{n-1}$, which we do in the sequel. 
Thus 
\[
\frac{\log 1/C}{D}\le  \log b + (n-1)\l\{\frac{\log (4D^3)}{D}+ \log (8 \cR b) +\log \|A\| \r\}
\le C_1(E)+C_2(E) \log \|A\|
\]  
where $C_1, C_2$ are constants which depend only on $E$.

\bigskip
In our application $A$ is equal to some $H_\omega^L$. Thus $ \|A\|=\|H_\omega^L\|\le \|H\| +\|q_0\|_\infty$ for all $\Lambda_L$ and $\omega$. Here $\|q_0\|_\infty$ denotes the essential supremum of the random variable $|q_0|$ when restricted to the set $\{\omega \mid |q_0(\omega)|< \infty\}$. Thus we obtain for the IDS of $H_\omega^L$ the estimate:
\[
N_\omega^L(E+\epsilon) - N_\omega^L(E) \le \frac{C_E }{\log(1/\epsilon)}
\]
and Theorem \ref{t-GlogHoelder} is proven.
\bigskip

We have established that the family of functions $N_\omega^\Lambda$, where $\Lambda$ runs through 
an admissible exhaustion, is equicontinuous from the right at the point $E$. 
To conclude Corollary \ref{c-GGlobConv} we will apply a simple lemma concerning the convergence of distribution functions. Together with the weak convergence 
of $\{N_\omega^\Lambda\}_\Lambda$ it will imply $N_\omega^\Lambda(E) \to N(E)$ as $\Lambda \to X$.

\begin{lem}
Let $N, N_L\colon \RR \to [0,\infty[$ for  $L \in \NN$  be monotone increasing, right-continuous functions, 
each of which is constant on the intervals $]-\infty, -M]$ and $[M, \infty[$ for some $L$-independent $M$.
Assume that for all $f\in C_0(\RR)$ we have $\lim_{L\to \infty} N_L(f)=N(f)$. Let the family $N_L$ be equicontinuous from the right at $E\in \RR$, i.e.~there exists a function $\delta$ such that $\lim_{\epsilon\searrow 0}\delta(\epsilon)=0$ and
\be
\label{e-StetModul}
N_L(E+\epsilon)-N_L(E)\le \delta(\epsilon) \text{ for all } L\in \NN .
\ee
Then $\lim_{L\to \infty} N_L(E)=N(E)$.
\end{lem}
\begin{proof}
For $\epsilon > 0$, let $f_\epsilon\in C_0(\RR)$ be a function with support in $[-M-1,E+\epsilon]$, such that $f_\epsilon(x)=1$ for all $x\in [-M, E]$, and $f_\epsilon$ is monotone  on $[E, E+\epsilon]$. By monotonicity and 
\eqref{e-StetModul} we have
\[
N_L(f_\epsilon)-\delta(\epsilon) \le  N_L(E) \le N_L(f_\epsilon) .
\]
First we send $L\to \infty$
\[
N(f_\epsilon) -\delta(\epsilon) \le \liminf_{L\to \infty} N_L(E) \le \limsup_{L\to \infty} N_L(E)
\le N(f_\epsilon) \le N(E+\epsilon)
\]
and then $\epsilon\to 0$ to obtain, using right continuity of $N$
\[
N(E) \le \liminf_{L\to \infty} N_L(E) \le \limsup_{L\to \infty} N_L(E) \le N(E) .
\]
 \end{proof}

The same arguments apply to the restrictions to the infinite cluster. 

\section{Estimates \`a la Wegner and Delyon-Souillard}
\label{s-Wegner}
In this section we prove Theorems \ref{t-GDelyonS} and \ref{t-GWegner}. The proofs hold for the infinite cluster operator $H_\omega^\infty$ as well.
We assume that the random variables $q_k$ are iid with distribution $\mu$. First we show that jumps of the IDS imply the existence of atoms of $\mu\upharpoonright_\RR$. To this end we use a lemma of Stollmann \cite{Stollmann-00b}:
\smallskip

\begin{lem}
\label{l-Stollmanns}
Let $\rho$ be a finite measure on $\RR $, $\Phi\colon\RR^J\to \RR$ continuous and monotone in each variable, such that 
\[
\Phi(q+t \, e) - \Phi(q) \ge t  \quad \text{ where } e = (1,\dots,1)\in \RR^J, t\in \RR .
\]
Set $s(\rho,\epsilon):=\sup\{\rho[a,a+\epsilon]\mid a \in \RR\}$. Then we have for any interval $I$
\[
(\otimes^J \rho)  \{q\mid \Phi(q)\in I \}\le J \, s(\rho, |I|) \, \rho(\RR)^{J-1} .
\]
\end{lem}
\smallskip

If $N$ is discontinuous at $E$, by Proposition \ref{p-equiv} there exists a finite $\Lambda\subset X$ and a set $\Omega'\subset \Omega$ of positive measure such that for all $\omega\in \Omega'$ exists a function $f=f_\omega$ with support in $\Lambda$ and $H_\omega f=Ef$. Since $f$ feels the potential only in $\Lambda^+:=\Lambda\cup \partial_R^o \Lambda$, for $\omega$ and $f$ as before we have $H_\omega^{\Lambda^+}f=Ef$. But $\PP\{\omega\mid H_\omega^{\Lambda^+}f=Ef\}>0$ implies that $\mu$ has an atom on $\RR$ as we will prove now:

Let $A$ be any subset of $\Lambda^+$. Set $\Omega(A):=\{\omega\mid q_j<\infty \forall j\in A, q_j=\infty \forall j\in \Lambda^+\setminus A\}$. Then $\Omega$ is the disjoint union of $\Omega(A), A\subset \Lambda^+$. Thus
\[
0<\PP\{\omega\mid H_\omega^{\Lambda^+}f=Ef\} = \sum_A \PP\{\omega\in \Omega(A)\mid H_\omega^{\Lambda^+}f=Ef\}.
\]
For $\omega\in \Omega(A)$ the operator $H_\omega^{\Lambda^+}=H_\omega^A$ acts on $\ell^2(A)$ and $\sum_{j\in A} \delta_j$ is the identity operator. Let $E_1(\omega), \dots,E_{|A|}(\omega)$ denote the eigenvalues of $H_\omega^A$. 
Then the function $\Phi(\omega):=E_n(\omega)$ is continuous and monotone increasing in each variable $q_j \in A$ and $\Phi(\omega+t \, e)-\Phi(\omega)=t$ where $t\in \RR$ and $e=(1,\dots, 1) \in \RR^A$. Note that if $\omega\in \Omega(A)$ then $\omega+t \,e \in \Omega(A)$ for all $t\in \RR$. 
Thus we can apply Lemma \ref{l-Stollmanns} with $\rho=\mu\upharpoonright_\RR$ and obtain 
\begin{align*}
\PP\{\omega\in \Omega(A)\mid E \in \sigma(H_\omega^{\Lambda^+})\}
& \le
\sum_n \PP\{\omega\in \Omega(A)\mid E_n(\omega)=E\}
\\
& \le 
\sum_n \PP\{\omega\in \Omega(A)\mid E_n(\omega)\in B_\epsilon(E)\}
\le \vert A\vert^2 \, s(\rho, \epsilon)
\end{align*}
for every $\epsilon>0$. If $\rho_{pp}$ is void, $ \inf_{\epsilon>0} s(\rho, \epsilon)=0$. This yields a contradiction and proves Theorem \ref{t-GDelyonS}. \qedsymbol

\medskip
Now we turn to the proof of the Wegner estimate (Theorem \ref{t-GWegner}). Thus we consider the situation where 
$\sigma(H)\subset [s_-,s_+]$ and $\tilde \mu:=\mu\upharpoonright_{]a+s_-, b+s_+[}$ has a density $f\in L^\infty$. 

Similarly as in the previous proof we partition the probability space according to the variables which lie in the singular, respectively absolutely continuous range of values. Let $A \subset \Lambda$ and 
\begin{align*}
\Lambda^{ac}(\omega)   &:= \{j \in \Lambda\mid q_j \in ]a+s_-, b+s_+[  \}     &&
\\
\Lambda^{sing}(\omega) &:= \{j \in \Lambda\mid q_j \not\in ]a+s_-, b+s_+[  \} && = \Lambda \setminus \Lambda^{ac}(\omega)
\\
\Omega(A)              & := \{\omega\in \Omega\mid \Lambda^{ac}(\omega)= A\}
& &= \{\omega\in \Omega\mid q_j\in ]a+s_-, b+s_+[ \Leftrightarrow j\in A\} .
\end{align*}
The family $\Omega(A), A \subset \Lambda$ forms a disjoint cover of $\Omega$.

\begin{lem} 
Fix $\delta>0$, let $I\subset \RR$ be such that $\dist(I,]a, b[^c)\ge \delta$, and assume that an eigenvalue $E_n(\omega)$ of $H_\omega^\Lambda$ is contained in $I$.
Then 
\[
\sum_{j \in \Lambda^{ac}(\omega)} \frac{\partial E_n(\omega)}{\partial q_j} 
\ge  \l (\frac{\delta}{b-a+s_+-s_-+1}\r)^2 .
\]
\end{lem}

\begin{proof}
First note that if $q_j$ is in $]a+s_-, b+s_+[$, so is a neighbourhood of $q_j$. 
Thus when defining the derivative $\frac{\partial }{\partial q_j}$ we can assume that $\Lambda^{ac}(\omega)$ is fixed.
We set $q_c =b+s_+ +1$ and partition the potential 
\[
V_\omega^{ac}= \sum_{j \in \Lambda^{ac}(\omega)} (q_j -q_c) \, \delta_j,
\quad 
V_\omega^{sing}= 
\sum_{j \in \Lambda^{sing}(\omega)} q_j \, \delta_j + \sum_{j \in \Lambda^{ac}(\omega)} q_c \, \delta_j .
\]
Let $\psi$ be a normalised eigenfunction corresponding to $E_n(\omega)$. Then 
\begin{align*}
\sum_{j \in \Lambda^{ac}(\omega)} \frac{\partial E_n(\omega)}{\partial q_j}
=
\sum_{j \in \Lambda^{ac}(\omega)}  \la \psi, \delta_j \psi \ra
&\ge 
\frac{\|V_\omega^{ac}\,\psi\|^2}{(b-a+s_+-s_-+1)^2}
\\
&=
\frac{\|(H^\Lambda+V_\omega^{sing}-E_n(\omega))\psi\|^2}{(b-a+s_+-s_-+1)^2} .
\end{align*}
Since $H^\Lambda+V_\omega^{sing}$ has no spectrum in $]a,b[$, we have 
$\|(H^\Lambda+V_\omega^{sing}-E_n(\omega))\psi\|^2 \ge \delta^2$ and the lemma is proven.
\end{proof}

Fix now $A\subset \Lambda$ and consider $\omega\in \Omega(A)$. To estimate 
$ \EE \{ \chi_{\Omega(A)} P_\omega^\Lambda(I)  \} $
we write $I^\circ=B_\epsilon(E)$ and proceed similarly as in \cite{Wegner-81,Kirsch-96}  and \cite{KirschV-02a}.  Introduce a smooth monotone function $\rho \colon \RR \to [0,1]$ 
taking the value 0 on $]-\infty, -\epsilon]$ and the value $1$ on $[\epsilon, \infty[$. 
Now 
\[
\chi_{]E-\epsilon,E+\epsilon [} (x)
\le 
\int_{-2\epsilon}^{2\epsilon} dt \,\rho'(x-E+t)
\]
implies by the spectral theorem
\[
\Tr \Big [ P_\omega^\Lambda(B_\epsilon(E)) \Big ]
\le
\Tr \Big [ \int_{-2\epsilon}^{2\epsilon} dt \,\rho'(H_\omega^\Lambda-E+t)  \Big ]
=
\sum_{n\in \NN} \int_{-2\epsilon}^{2\epsilon} dt \,\rho'(E_n^\Lambda(\omega)-E+t) .
\]
The chain rule implies
\begin{eqnarray*}
\label{e-HFlowerbound}
\rho'(E_n(H_\omega^\Lambda) -E + t)
\le
\l (\frac{b-a+s_+-s_-+1}{\delta}\r)^2  \sum_{j \in A} \frac{ \partial \rho (E_n(H_\omega^\Lambda) -E + t) }{ \partial q_j } 
\end{eqnarray*}
Since $\delta_j$ is a rank one operator, we have
\[
\int_{a+s_-}^{b+s_+}  dq_j \, f(q_j) \, 
\sum_{n\in \NN}\frac{\partial  \rho(E_n(H_\omega^\Lambda)-E+t)}{\partial q_j}
\le \|f\|_\infty 
\]
which implies
\[
\EE \l \{ \chi_{\Omega(A)} \sum_{n\in \NN} \frac{\partial  \rho(E_n(H_\omega^\Lambda)-E+t)}{\partial q_j} \r \}
\le \|f\|_\infty \  \frac{\EE \{ \chi_{\Omega(A)} \}}{\mu (]a+s_-, b+s_+ [)}.
\]
Therefore
\[
\EE \{ \chi_{\Omega(A)} P_\omega^\Lambda(I)\} 
\le 
4\l(\frac{b-a+s_+-s_-+1}{\delta}\r)^2   \|f\|_\infty\frac{\EE \{ \chi_{\Omega(A)} \} }{\mu (]a+s_-, b+s_+ [)} \ |A| \, \epsilon .
\]
Summing over $A\subset \Lambda$ gives the desired estimate \eqref{e-GWE}.

%

\begin{thebibliography}{dGLM59b}

\bibitem[Ada93]{Adachi-1993}
T.~Adachi.
\newblock A note on the {F\o lner} condition for amenability.
\newblock {\em Nagoya Math. J.}, 131:67--74, 1993.

\bibitem[And58]{Anderson-58}
P.W. Anderson.
\newblock Absence of diffusion in certain random lattices.
\newblock {\em Phys. Rev.}, 109:1492, 1958.

\bibitem[AS93]{AdachiS-1993}
T.~Adachi and T.~Sunada.
\newblock Density of states in spectral geometry.
\newblock {\em Comment. Math. Helv.}, 68(3):480--493, 1993.

\bibitem[CCF{\etalchar{+}}86]{ChayesCFST-86}
J.~T. Chayes, L.~Chayes, J.~R. Franz, J.~P. Sethna, and S.~A. Trugman.
\newblock On the density of states for the quantum percolation problem.
\newblock {\em J. Phys. A}, 19(18):L1173--L1177, 1986.

\bibitem[CHKN02]{CombesHKN-02}
J.-M. Combes, P.~D. Hislop, F.~Klopp, and S.~Nakamura.
\newblock The {W}egner estimate and the integrated density of states for some
  random operators.
\newblock {\em Proc. Indian Acad. Sci. Math. Sci.}, 112(1):31--53, 2002.
\newblock www.ias.ac.in/mathsci/.

\bibitem[CL90]{CarmonaL-90}
R.~Carmona and J.~Lacroix.
\newblock {\em Spectral Theory of Random {Schr\"odinger} Operators}.
\newblock Birkh\"auser, Boston, 1990.

\bibitem[CS83]{CraigS-83a}
W.~Craig and B.~Simon.
\newblock Log {H\"older} continuity of the integrated density of states for
  stochastic {Jacobi} matrices.
\newblock {\em Commun. Math. Phys.}, 90:207--218, 1983.

\bibitem[dGLM59a]{deGennesLM-59b}
P.-G. de~Gennes, P.~Lafore, and J.~Millot.
\newblock Amas accidentels dans les solutions solides d\'esordonn\'ees.
\newblock {\em J. of Phys. and Chem. of Solids}, 11(1--2):105--110, 1959.

\bibitem[dGLM59b]{deGennesLM-59a}
P.-G. de~Gennes, P.~Lafore, and J.~Millot.
\newblock Sur un ph\'enom\`ene de propagation dans un milieu d\'esordonn\'e.
\newblock {\em J. Phys. Rad.}, 20:624, 1959.

\bibitem[DLM{\etalchar{+}}03]{DodziukLMSY-03}
J.~Dodziuk, P.~Linnell, V.~Mathai, T.~Schick, and S.~Yates.
\newblock Approximating {$L^2$}-invariants, and the {Atiyah} conjecture.
\newblock {\em Comm. Pure Appl. Math.}, 56(7):839 -- 873, 2003.

\bibitem[DMY]{DodziukMY}
J.~Dodziuk, V.~Mathai, and S.~Yates.
\newblock Approximating ${L}^2$ torsion on amenable covering spaces.
\newblock math.DG/0008211 on arxiv.org, see also \cite{DodziukLMSY-03}.

\bibitem[DS84]{DelyonS-84}
F.~Delyon and B.~Souillard.
\newblock Remark on the continuity of the density of states of ergodic
  finite-difference operators.
\newblock {\em Commun. Math. Phys.}, 94:289--291, 1984.

\bibitem[Far98]{Farber-98}
M.~Farber.
\newblock Geometry of growth: approximation theorems for {$L\sp 2$} invariants.
\newblock {\em Math. Ann.}, 311(2):335--375, 1998.

\bibitem[Jes92]{Jeske-92}
F.~Jeske.
\newblock {\em \"Uber lokale Positivit\"at der Zustandsdichte zuf\"alliger
  Schr\"odinger-Operatoren}.
\newblock doctoral thesis, Ruhr-Universit\"at Bochum, 44801 Bochum, 1992.

\bibitem[KB97]{KantelhardtB-97}
J.~W. Kantelhardt and A.~Bunde.
\newblock Electrons and fractons on percolation structures at criticality:
  Sublocalization and superlocalization.
\newblock {\em Phys. Rev. E}, 56:6693–--6701, 1997.

\bibitem[KB98a]{KantelhardtB-98b}
J.~W. Kantelhardt and A.~Bunde.
\newblock Extended fractons and localized phonons on percolation clusters.
\newblock {\em Phys. Rev. Lett.}, 81:4907–--4910, 1998.

\bibitem[KB98b]{KantelhardtB-98}
J.~W. Kantelhardt and A.~Bunde.
\newblock Wave functions in the {A}nderson model and in the quantum percolation
  model: a comparison.
\newblock {\em Ann. Phys. (8)}, 7(5-6):400--405, 1998.

\bibitem[KB02]{KantelhardtB-02}
J.~W. Kantelhardt and A.~Bunde.
\newblock Sublocalization, superlocalization, and violation of standard
  single-parameter scaling in the {Anderson} model.
\newblock {\em Phys. Rev. B}, 66, 2002.

\bibitem[KE72]{KirkpatrickE-72}
S.~Kirkpatrick and T.~P. Eggarter.
\newblock Localized states of a binary alloy.
\newblock {\em Phys. Rev. B}, 6:3598, 1972.

\bibitem[Kir96]{Kirsch-96}
W.~Kirsch.
\newblock {Wegner} estimates and {Anderson} localization for alloy-type
  potentials.
\newblock {\em Math. Z.}, 221:507--512, 1996.

\bibitem[KLS03]{KlassertLS-03}
S.~Klassert, D.~Lenz, and P.~Stollmann.
\newblock Discontinuities of the integrated density of states for random
  operators on {D}elone sets.
\newblock {\em Comm. Math. Phys.}, 241(2-3):235--243, 2003.
\newblock arXiv.org/math-ph/0208027.

\bibitem[KM]{KirschM-04}
W.~Kirsch and P.~M\"uller.
\newblock Spectral properties of the laplacian on bond-percolation graphs.
\newblock math-ph/0407047 on arXiv.org.

\bibitem[KS04]{KostrykinS-04}
V.~Kostrykin and R.~Schrader.
\newblock A random necklace model.
\newblock {\em Waves in Random Media}, 14:S75 -- S90, 2004.
\newblock arxiv.org/math-ph/0309032.

\bibitem[KV02]{KirschV-02a}
W.~Kirsch and I.~Veseli{\'c}.
\newblock Wegner estimate for sparse and other generalized alloy type
  potentials.
\newblock {\em Proc. Indian Acad. Sci. Math. Sci.}, 112(1):131--146, 2002.
\newblock www.ias.ac.in/mathsci/, mp\_arc-bin/02-143.

\bibitem[Lin01]{Lindenstrauss-01}
E.~Lindenstrauss.
\newblock Pointwise theorems for amenable groups.
\newblock {\em Invent. Math.}, 146(2):259--295, 2001.

\bibitem[LPV04]{LenzPV-04}
D.~Lenz, N.~Peyerimhoff, and I.~Veseli{\'c}.
\newblock Integrated density of states for random metrics on manifolds.
\newblock {\em Proc. London Math. Soc. (3)}, 88(3):733--752, 2004.

\bibitem[L{\"u}c94]{Lueck-94c}
W.~L{\"u}ck.
\newblock Approximating {$L\sp 2$}-invariants by their finite-dimensional
  analogues.
\newblock {\em Geom. Funct. Anal.}, 4(4):455--481, 1994.

\bibitem[L{\"u}c02]{Lueck-02}
W.~L{\"u}ck.
\newblock {\em {$L\sp 2$}-invariants: theory and applications to geometry and
  {$K$}-theory}, volume~44 of {\em Ergebnisse der Mathematik und ihrer
  Grenzgebiete. 3rd Series.}
\newblock Springer-Verlag, Berlin, 2002.

\bibitem[MSY03]{MathaiSY-03}
V.~Mathai, T.~Schick, and S.~Yates.
\newblock Approximating spectral invariants of {H}arper operators on graphs.
  {II}.
\newblock {\em Proc. Amer. Math. Soc.}, 131(6):1917--1923 (electronic), 2003.

\bibitem[MY02]{MathaiY-02}
V.~Mathai and S.~Yates.
\newblock Approximating spectral invariants of {H}arper operators on graphs.
\newblock {\em J. Funct. Anal.}, 188(1):111--136, 2002.
\newblock arXiv.org/math.FA/0006138.

\bibitem[PF92]{PasturF-92}
L.~A. Pastur and A.~L. Figotin.
\newblock {\em Spectra of Random and Almost-Periodic Operators}.
\newblock Springer Verlag, Berlin, 1992.

\bibitem[PV02]{PeyerimhoffV-02}
N.~Peyerimhoff and I.~Veseli\'c.
\newblock Integrated density of states for ergodic random {Schr\"{o}dinger}
  operators on manifolds.
\newblock {\em Geom. Dedicata}, 91(1):117--135, 2002.

\bibitem[SAH82]{ShapirAH-82}
Y.~Shapir, A.~Aharony, and A.~B. Harris.
\newblock Localization and quantum percolation.
\newblock {\em Phys. Rev. Lett.}, 49(7):486--489, 1982.

\bibitem[Sto00]{Stollmann-00b}
P.~Stollmann.
\newblock Wegner estimates and localization for continuum {A}nderson models
  with some singular distributions.
\newblock {\em Arch. Math. (Basel)}, 75(4):307--311, 2000.

\bibitem[Sto01]{Stollmann-01}
P.~Stollmann.
\newblock {\em Caught by disorder: A Course on Bound States in Random Media},
  volume~20 of {\em Progress in Mathematical Physics}.
\newblock Birkh\"auser, 2001.

\bibitem[Ves]{Veselic-QP}
I.~Veseli\'c.
\newblock Quantum site percolation on amenable graphs.
\newblock to appear in \emph{Applied Mathematics and Scientific Computing},
  June 2003, Brijuni, arXiv.org/math-ph/0308041.

\bibitem[Ves04]{Veselic-04a}
I.~Veseli\'c.
\newblock Integrated density of states and {Wegner} estimates for random
  {Schr\"odinger} operators.
\newblock In R.~del Rio and C.~Villegas-Blas, editors, {\em Schr\"odinger
  operators (Universidad Nacional Autonoma de Mexico, 2001)}, volume 340 of
  {\em Contemp. Math.}, pages 98--184. Amer. Math. Soc., Providence, RI, 2004.
\newblock arXiv.org/math-ph/0307062.

\bibitem[Weg81]{Wegner-81}
F.~Wegner.
\newblock Bounds on the {DOS} in disordered systems.
\newblock {\em Z. Phys. B}, 44:9--15, 1981.

\end{thebibliography}
\newcommand{\etalchar}[1]{$^{#1}$}
\def\cprime{$'$}

\end{document}